\documentstyle[11pt,newpasp,twoside,psfig]{article}
\def\edcomment#1{\iffalse\marginpar{\raggedright\sl#1\/}\else\relax\fi}
\reversemarginpar

\newcommand\HII{H\,{\sc ii}}

\def\etal{{\rm et~al.}}

\begin{document}
\title{Radio Emission from the Milky Way}
\author{Bryan Gaensler}
\affil{Harvard-Smithsonian Center for Astrophysics, 60 Garden Street MS-6,
Cambridge MA 02138, USA}

\begin{abstract}

In the last 5--10 years, wide-field imaging capabilities and
effective mosaicing algorithms have made possible a variety of
ambitious interferometric surveys of the Galaxy, resulting in images
of unprecedented sensitivity and resolution. Here I discuss some of
the highlights from these new surveys. Amongst the new results are
the identification of many new supernova remnants in confused regions,
spectacular low frequency images of the inner Galaxy with the VLA, and
some remarkable new insights into the structure of the Galactic magnetic
field from linear polarization.

\end{abstract}

\section{Historical Overview}

In the 1930s and 1940s, the pioneering efforts of Jansky and Reber
demonstrated that the disk of the Milky Way was a strong source of
radio emission. In the 1950s, a theoretical framework was developed
which argued that this emission was due to the synchrotron process ---
this was soon confirmed when it was demonstrated that this emission was
linearly polarized. Over the next few decades, various groups in both
hemispheres proceeded to survey the radio emission from the Milky Way
using both single-dish and pencil beam surveys, an example of which is
shown in Figure~1. These surveys demonstrated that the radio
emission from the Milky Way was comprised of three main components:
thermal emission from individual \HII\ regions, synchrotron emission
from discrete supernova remnants (SNRs), all  superimposed on diffuse
synchrotron emission from the relativistic interstellar medium (ISM).

While these surveys provided a very useful basis for identifying
and classifying many Galactic sources, the lack of sensitivity and
resolution in these data leaves serious deficiencies in our understanding
of the populations of radio sources in our Galaxy.  Specifically, since
SNRs and \HII\ both trace massive star formation, these sources tend
to be clustered in the same parts of the sky. This makes it difficult
to identify both small-diameter (i.e.\ young) and faint (i.e.\ old)
SNRs, important for understanding SNR birthrates and SNR lifetimes,
respectively.  Many sources are too compact to be identified as SNRs,
\HII\ regions or background sources on a morphological basis alone, and
there are too many of these sources to study each object individually at
higher resolution and at other wavelengths. Many interesting objects have
thus been overlooked, and only later identified serendipitously (e.g.\
Duncan \& Green 2000; Crawford \etal\ 2001); no doubt many more such
objects are still to be found.  The net result is that studies based
on the Galactic plane surveys of the 1970s and 1980s have resulted in
incomplete catalogues, biased demographics and many overlooked objects
(e.g.\ Helfand \etal\ 1989; Green 1991).

\begin{figure}
%\centerline{\psfig{file=fig_green.eps,width=\textwidth}}
\centerline{\psfig{file=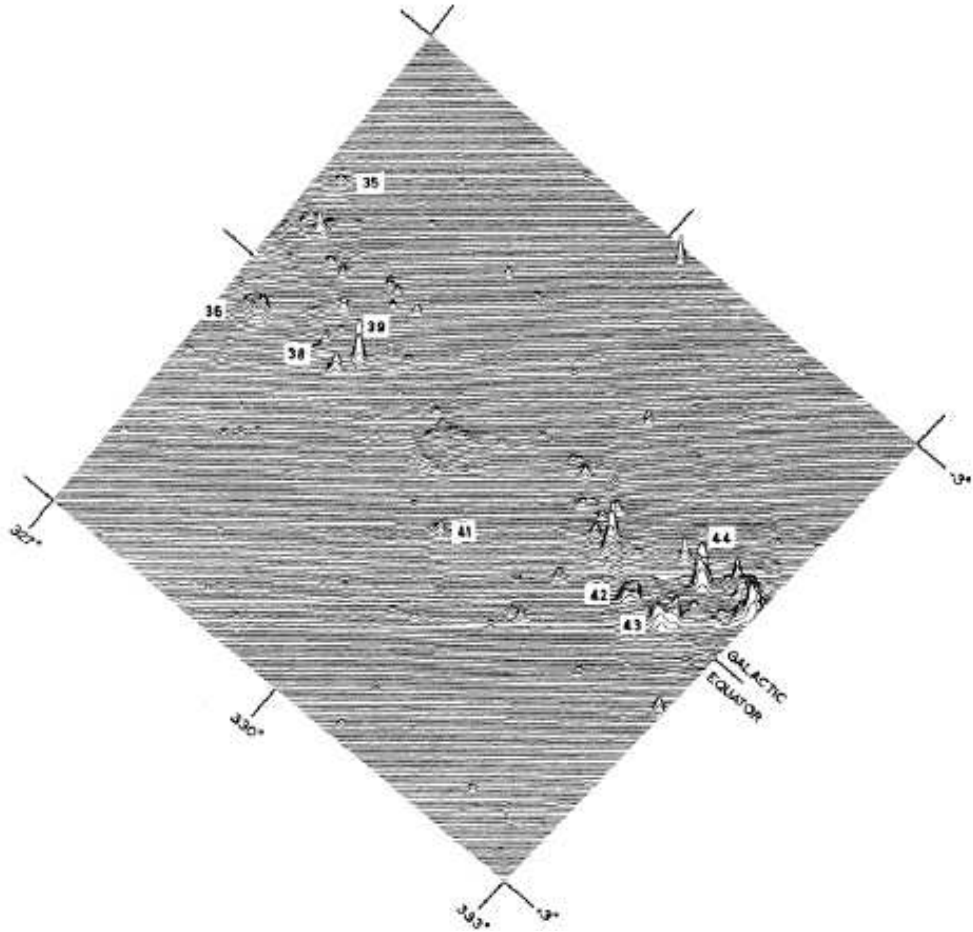,width=\textwidth}}
\label{fig_green}
\caption{Part of the Mills Cross Galactic plane survey, at a frequency
of 408 MHz and a spatial resolution of 3~arcmin (Green 1974).}
\end{figure}

In the rest of this paper, I will discuss the ways in which new surveys
are overcoming the limitations of these earlier efforts.

\section{Synthesis Imaging Arrays}

The most important advance in studying Galactic radio emission is
that there are now high-fidelity synthesis imaging telescopes in both
hemispheres. Most of these instruments have now carried out sensitive
surveys of the Galactic plane. The immediate advantage of these
observations is that one can image the full field of view of a single
antenna element, but can obtain the spatial resolution of the longest
baseline and the sensitivity corresponding to the total collecting
area. Many interferometric surveys map large regions of the sky simply
by imaging individual fields, and then ``pasting'' these fields together.

One of the most extensive such surveys was the Molonglo Galactic Plane
Survey (MGPS) (Figure~2; Green \etal\ 1999), consisting of
an 843 MHz image at a spatial resolution of $43''$, covering 330~deg$^2$
of the southern Galaxy. While morphology alone does not always distinguish
\HII\ regions from SNRs in such data, a simple comparison with {\em IRAS}\
images allows one to clearly differentiate between \HII\ regions (which
are generally prominent infrared sources) and SNRs (which are usually
not detected in {\em IRAS}\ data). In the MGPS, this process enabled
the identification of 75 SNRs, 25\% of which were newly discovered
(Whiteoak \& Green 1996).  The VLA continuum survey of Becker \etal\
(these proceedings) is employing a similar approach at higher spatial
resolution.

\begin{figure}
%\centerline{\psfig{file=fig_mgps_xfig.eps,width=\textwidth,clip=}}
\centerline{\psfig{file=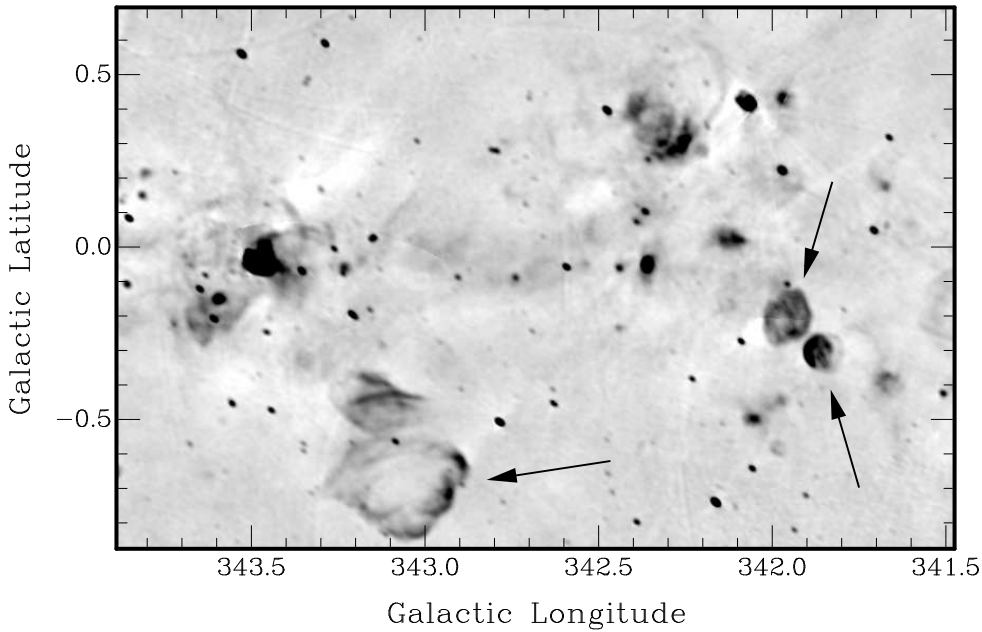,width=\textwidth,clip=}}
\caption{A region from the 843 MHz MGPS, showing various SNRs, \HII\ regions
and background sources at sub-arcmin resolution (Green \etal\ 1999).
The arrows mark the three known SNRs in this field. These SNRs
are conspicuous for their lack of infrared emission,
in contrast with the \HII\ regions.}
\label{fig_mgps}
\end{figure}

\section{Single Dish Combination}

While the longest baseline of an interferometric array sets the smallest
scale detectable in the image, it is often not appreciated that the
{\em shortest}\ baseline of the array sets the {\em largest}\ scale to
which the image is sensitive.  An image made from interferometer data
alone will therefore miss large-scale structure, and the resulting flux
densities and spectral index determinations will be incorrect. The
only way to correct this problem is to map the same region with a
single dish. As shown in Figure~3, the single-dish and
interferometric data can then be combined to provide a complete picture
of the distribution of radio emission on the sky.

\begin{figure}
%\centerline{\psfig{file=fig_single_dish_comb.eps,width=\textwidth}}
\centerline{\psfig{file=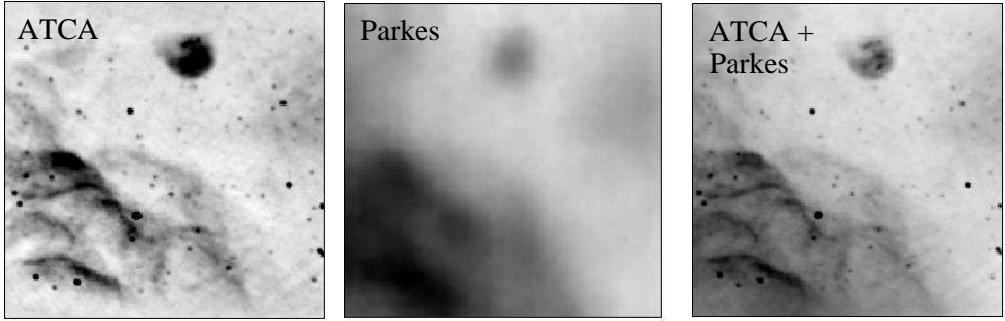,width=\textwidth}}
\caption{1.4 GHz continuum images from the Southern Galactic Plane survey
(SGPS),
showing a small region of the southern Galaxy encompassing the \HII\
region RCW~32 and part of the Vela SNR (McClure-Griffiths 2001). The
ATCA data are sensitive to high spatial resolution features, while the
Parkes data detect large-scale structure. Only in the combined image do
we see the true sky at arcmin resolution.}
\label{fig_comb}
\end{figure}

The most important demonstration of this process has been the Canadian
Galactic Plane Survey (CGPS; Taylor \etal\ 2003).  The CGPS team have
gone to great efforts to ensure that at both 408 and 1420~MHz their
synthesis data are properly combined with single-dish maps. With such
data, SNRs and \HII\ regions can be clearly distinguished on the basis of
their spectral indices alone, while with the proper imaging of structure
on all scales, the relationship between individual objects and diffuse
emission becomes clearly apparent.

\section{Wide Field Imaging}

A key limitation in surveys carried out with synthesis telescopes is
the efficiency with which such surveys can cover large parts of the
sky --- both the MGPS and SGPS took many years to complete. However,
several recent developments have greatly enhanced the efficiency and
feasibility of such efforts.

First, some interferometers (most notably the ATCA) employ very rapid
source switching, which minimises overheads and enables large mosaics. For
example, the test region of the SGPS (McClure-Griffiths \etal\ 2001)
consisted of 190 pointings, 40 snapshots per pointing and 30 seconds per
snapshot, resulting in arcmin resolution images of the Galactic plane at a
mapping speed of 0.5~deg$^2$ per hour.  Second, many arrays now offer very
compact array configurations. This allows the $u-v$ plane to be quickly
filled, minimising the number of snapshots needed per pointing.  Finally,
rather than covering large areas of the sky just by pasting together
individual fields, new algorithms utilise the extra information contained
in adjacent pointings, which greatly improves the $u-v$ coverage of the
observation (Ekers \& Rots 1979). As demonstrated by Cornwell (1988),
this technique can be realised through joint deconvolution techniques,
resulting in mosasiced images of far greater fidelity than maps generated
by the ``individual'' approach.

Since the linear diameter of the field of view is proportional to the
observing wavelength, another way to map very large areas of the sky is to
carry out synthesis imaging at low frequencies.  For example, the 74-MHz
system at the VLA provides a field of view of $\sim70$~deg$^2$ per
pointing! Despite this obvious advantage, radio frequency interference,
ionospheric distortion of the wavefront and the non-coplanarity
of the VLA all make low-frequency imaging extremely challenging. A
variety of concerted efforts have now addressed these issues, allowing
spectacular wide-field images to be produced of the Galactic Centre and
other complicated regions (e.g.\ LaRosa \etal\ 2000). This has given us
the confidence to embark on a more ambitious 74/327 MHz Galactic plane
survey with the VLA, which will ultimately cover the range $-15^\circ <
l < 55^\circ$ (e.g.\ Brogan \etal\ 2001). 
Such a project should yield the identification of
many new low surface brightness and steep spectrum sources, such as SNRs,
pulsars, and high-redshift radio galaxies.

\section{Polarimetry}

The advent of high polarization purity, flexible correlators, and new
algorithms have made it feasible to produce wide-field images of Stokes Q,
U and V (e.g.\ Gaensler \etal\ 2001; Uyan{\i}ker \etal\ 2003). As shown
in Figure~4, these images can appear completely different
from the distribution of emission seen in Stokes~I.

\begin{figure}
%\centerline{\psfig{file=fig_poln.eps,width=\textwidth,angle=270,clip=}}
%\centerline{\psfig{file=gaensler_b_4.eps,width=\textwidth,angle=270,clip=}}
\vspace{5cm}
\caption{1.4 GHz continuum images of the SGPS test region (Gaensler \etal\
2001). The upper panel shows the total intensity emission from the region,
which originates from the usual mix of SNRs, \HII\ regions and background
sources. The lower panel shows linearly polarized intensity from the
same region.  Little correspondence is seen between the two panels.
(Note that the maximum brightness seen in linear polarization is
approximately ten times fainter than that seen in total intensity.)}
\label{fig_poln}
\end{figure}

Not surprisingly, these images detect linear polarization from
individual SNRs and from background point sources, the Faraday
rotation towards which can be measured and can be used to map out the
structure of the Galactic magnetic field (e.g.\ Brown \& Taylor 2001).
However, in addition to these features, we also see copious diffuse
polarization with no counterpart in total intensity, produced by
foreground Faraday rotation of the diffuse synchrotron background by
warm ionized gas (Figure~4; Uyan{\i}ker \etal\ 2003).  Many regions also
show significant depolarization, resulting from small-scale changes in
polarization position angle produced by turbulent foreground regions
(Gaensler \etal\ 2001). Finally, several groups now see polarimetric
structures corresponding to as yet unidentified magnetoionic structures
(Gray \etal\ 1998).  These polarimetric studies are thus enabling an
entirely new way of viewing the ISM.

\section{Conclusions}

A variety of new techniques are now being employed in surveys of
the radio emission from the Milky Way. These include:

\begin{itemize}
\item Synthesis imaging surveys, which greatly improve on the
resolution and sensitivity of earlier efforts.
\item Single-dish combination, which improves the spatial dynamic
range and provides sensitivity to diffuse structure.
\item Mosaicing and low frequency imaging, which generate higher fidelity
images and wider fields of view.
\item Polarimetric imaging, which is providing a whole new view of
the magnetic Milky Way.
\end{itemize}
Only with the surveys which are consequently emerging are we finally
now obtaining a complete picture of Galactic radio emission and its
constituent components.

\acknowledgements

The Australia Telescope is funded by the Commonwealth of Australia for
operation as a National Facility managed by CSIRO. I acknowledge
the support of the National Science Foundation through grant AST-0307358.

\end{document}